\documentclass[sigconf]{acmart}

\setcopyright{none}
\renewcommand\footnotetextcopyrightpermission[1]{}

\usepackage{fancyhdr}
\fancypagestyle{plain}{%
   \fancyhf{} %
   \fancyfoot[L]{ACM SIGCOMM Computer Communication Review }%
   \fancyfoot[R]{Volume 55 Issue 3, July 2025}%
   
}
\usepackage{balance}
\usepackage{subfigure}
\usepackage{booktabs}
\usepackage{multirow}
\usepackage{array}
\usepackage{xurl}
\usepackage{xspace}

\newcommand{\testbed}{\textsc{Campus5G}\xspace}
\newcommand{\mypar}[1]{\noindent\textbf{#1.}\xspace}
\newenvironment{tightlist}{
\begin{list}{$\bullet$}{
    \setlength{\topsep}{.1em}
    \setlength{\partopsep}{0in}
    \setlength{\parskip}{0in}
    \setlength{\itemsep}{0in}
    \setlength{\parsep}{0in}
    \setlength{\leftmargin}{1em}
    \setlength{\rightmargin}{0in}
    \setlength{\itemindent}{0in}
}}
{\end{list}}

\usepackage{titlesec}
\titlespacing{\section}{0pt}{0.3em}{0.3em}
\titlespacing{\subsection}{0pt}{0.3em}{0.3em}

\usepackage[skip=2pt]{caption}

\begin{document}

\title{Campus5G: A Campus Scale Private 5G Open RAN Testbed}
\author{Andrew E. Ferguson}
\authornote{Equal contributors to this article.}
\orcid{0009-0005-5331-1181}
\affiliation{%
  \institution{The University of Edinburgh}
  \country{United Kingdom}
}
\email{Andrew.E.Ferguson@ed.ac.uk}

\author{Ujjwal Pawar}
\authornotemark[1]
\orcid{0000-0002-5719-3427}
\affiliation{%
  \institution{The University of Edinburgh}
  \country{United Kingdom}
}
\email{u.pawar@sms.ed.ac.uk}

\author{Tianxin Wang}
\affiliation{%
  \institution{The University of Edinburgh}
  \country{United Kingdom}
}
\email{twang4@ed.ac.uk}

\author{Mahesh K. Marina}
\affiliation{%
  \institution{The University of Edinburgh}
  \country{United Kingdom}
}
\email{mahesh@ed.ac.uk}

\begin{abstract}
	Mobile networks are embracing disaggregation, reflected by the industry trend towards Open RAN. 
    Private 5G networks are viewed as particularly suitable contenders as early adopters of Open RAN, owing to their setting, high degree of control, and opportunity for innovation they present. 
    Motivated by this, we have recently deployed \testbed, the first of its kind campus-wide, O-RAN-compliant private 5G testbed across the central campus of the University of Edinburgh. 
    We present in detail our process developing the testbed, from planning, to architecting, to deployment, and measuring the testbed performance. 
    We then discuss the lessons learned from building the testbed, and highlight some research opportunities that emerged from our deployment experience. 
\end{abstract}


\maketitle

\section{Introduction}
\label{sec:intro}

Mobile networks in the recent past have embraced {\em disaggregation} at different levels, driven by the need for cost-effective and flexible system architecture that accelerates innovation and enables new services.
Complimentarily, they are headed towards data-driven and AI powered operation to unlock significant energy, operational and spectral efficiency gains. 
These trends are most prominently evident with {\em Open Radio Access Network (RAN)}~\cite{open-ran-overview}, being standardized by the O-RAN Alliance~\cite{oran-alliance}.

The O-RAN architecture~\cite{oran-architecture} disaggregates the traditional monolithic base stations into Centralized (CU), Distributed Unit (DU) and Radio Unit (RU) components that communicate over open standardized interfaces. 
Of these components, CU and DU are intended to be realized as virtualized network functions (VNFs) over commodity compute hardware, thus reduce CAPEX for operators. 
This disaggregated architecture also promises to diversify the mobile telecoms ecosystem, as it offers flexibility for operators to mix and match solutions from different vendors. 
In addition, the O-RAN architecture features RAN Intelligent Controllers (RICs) operating at different timescales and with corresponding set of ``Apps'' (for energy saving, load balancing, anomaly detection, etc.) that can leverage AI/ML towards efficient RAN operation. 

While Open RAN is expected to become an integral part of future (6G) mobile network system architecture generally~\cite{polese-jsac24}, its deployment in national scale public mobile networks today is mostly limited to trials~\cite{oran-map,open-ran-current-state}.
This could be partly attributed to Open RAN not making it to the last investment cycle for public mobile network operators (MNOs) when they transitioned to 5G~\cite{open-ran-outlook}. 

It is in this context that {\em private 5G mobile networks}~\cite{private-5g} are seen as the early adopters of Open RAN. 
Unlike national scale public mobile networks, private networks are ``local'' scale deployments that target industrial (e.g., manufacturing, mining, warehouses) and enterprise (e.g., business parks, universities, hospitals) settings, transport hubs, venues and such~\cite{analysys-mason-report}. 
Private mobile networks are appealing for these environments because they not only allow high degree of control and customization but also offer greater opportunity for innovation (e.g., edge AI applications~\cite{ericsson-mobility-report,foukas-hotmobile25,emdl22}). 
Private 5G deployments are rapidly growing~\cite{private5g-european-5g-observatory,bmw,bosch,arana,private-network-deployments-2023,uk-private-5g-projects}.
Private 5G is a key use case for Open RAN~\cite{martian25}. 
New breed of operators and vendors target private 5G with Open RAN as a default~\cite{telet,neutral-wireless,jet-connectivity,accelleran}. 

University campuses are a representative and compelling setting for Open RAN based private 5G networks. Moreover, as history has shown with software-defined networking (SDN), campus networks can be excellent experimentation grounds for innovative technological solutions and services that have the potential to shape our future connectivity~\cite{openflow2008}.

Motivated by the above, we have recently deployed \testbed, the first of its kind campus-wide, O-RAN-compliant private 5G testbed across the central campus of the University of Edinburgh. 
Our deployment setting is a typical dense urban outdoor environment with high footfall and mobile traffic demand. 
We designed \testbed~ with blanket coverage across the campus in mind while at the same time allowing ample flexibility for experimentally evaluating innovative designs across the whole O-RAN architecture in a real-world setting.
This includes the ability to use commercial and open source software components from different vendors in the Open RAN ecosystem as well as research prototypes. 
We also embed AI compute to facilitate research on AI for RAN and edge AI applications. 
From a user perspective, our testbed allows the use of unmodified commodity mobile devices through custom physical or eSIMs for our network. 

\testbed~ is intended to enable broad ranging experimental research on next generation wireless access networks,
including experimentally exploring systems research challenges in the Open RAN context;
developing and validating innovative data-driven solutions for optimized RAN operation and beyond; and
investigating issues unique to private 5G networks and their role in the future mobile network system architecture.
In the longer term, we envision this testbed to become a living lab routinely used by university staff and students.

In this paper, we present our end to end process with the \testbed~testbed deployment from the planning stage and architecture design to component selection, configuration and measurement based coverage analysis. 
Compared to existing private 5G or Open RAN testbeds (discussed in \S\ref{sec:related-work} and summarized in Table~\ref{tab:testbed-comparison}), our testbed is unique not only in its campus level scale, high degree of flexibility and fully O-RAN compliant nature but also due to its deployment in an urban outdoor environment. 
We share some of the key lessons we learned from the testbed deployment as well as potential research opportunities we identified in the process.
We hope this paper can serve as a blueprint to inform other researchers intending to deploy similar testbeds. 
Equally, we believe our experience reported in this paper will benefit Open RAN research and experimentation in the context of private 5G and beyond.

\section{Related Work}
\label{sec:related-work}

By way of positioning our \testbed~ testbed with respect to existing private 5G and Open RAN testbeds, here we briefly discuss those prior efforts. 
There exist a few private 5G deployments made up of commercial components that are being used as testbeds. 
Examples include NICT Local 5G testbed in Japan~\cite{nict-local5g}, South Bend CBRS network in the US~\cite{south-bend} and Fraunhofer IIS Industrial Testbed in Germany~\cite{5g-opera}. 
Some of these like NICT and South Bend testbeds are not O-RAN compatible while others like the Fraunhofer testbed are primarily meant for interoperability testing and so offer very limited flexibility for research use.
\begin{table}[t]
    \centering
        \captionsetup{justification=centering}
    \caption{\testbed~ versus representative set of existing private 5G Open RAN testbeds.}
    \resizebox{\linewidth}{!}
    {
        \begin{tabular}{|l|
        >{\columncolor[HTML]{FFFC9E}}l|
        >{\columncolor[HTML]{96FFFB}}l|l|l|l|}
        \hline
        \textbf{Testbed} & \textbf{Size} & \textbf{\begin{tabular}[c]{@{}l@{}}Indoor or\\ Outdoor\end{tabular}} & \textbf{\begin{tabular}[c]{@{}l@{}}O-RAN\\ Compatibility\end{tabular}} & \textbf{\begin{tabular}[c]{@{}l@{}}Commercial\\ RUs or SDRs\end{tabular}} & \textbf{\begin{tabular}[c]{@{}l@{}}Software\\ Flexibility\end{tabular}} \\ \hline
        Arena~\cite{arena} & 24 & Indoor & Partially & SDRs & Open Source \\ \hline
        Powder~\cite{powder} & 8 & Outdoor & Partially & SDRs & Open Source \\ \hline
        Colosseum~\cite{colosseum} & 128 & Emulated & Partially & SDRs & Open Source \\ \hline
        X5G~\cite{x5g} & 8 & Indoor & Yes & Commercial RUs & \begin{tabular}[c]{@{}l@{}}Open Source\\ or Commercial\end{tabular} \\ \hline
        \begin{tabular}[c]{@{}l@{}}Microsoft\\ Cambridge~\cite{microsoft-testbed}\end{tabular} & 20 & Indoor & Yes & Commercial RUs & \begin{tabular}[c]{@{}l@{}}Open Source\\ or Commercial\end{tabular} \\ \hline
        SIT-FCT~\cite{sit-fct-oran} & 5 & \begin{tabular}[c]{@{}l@{}}Indoor and\\Outdoor\end{tabular} & Yes & Commercial RUs & Commercial \\ \hline
        OpenRAN@Brazil~\cite{brasil} & 8 & \begin{tabular}[c]{@{}l@{}}Indoor and\\Outdoor\end{tabular} & Yes & Commercial RUs & Commercial \\ \hline
        \textbf{\testbed} & \textbf{20} & \textbf{Outdoor} & \textbf{Yes} & \textbf{Commercial RUs} & \textbf{\begin{tabular}[c]{@{}l@{}}Open Source\\ or Commercial\end{tabular}} \\ \hline
        \end{tabular}
    }

    \label{tab:testbed-comparison}
    \vspace{-0.5cm}
\end{table}

Many testbeds currently exist to support O-RAN research. 
These testbeds by their nature are private (4G/5G) networks.
Majority of them rely on USRP software defined radio (SDR) based radio frontends for the RUs, and are for the most part deployed indoors. 
Arena testbed~\cite{arena} is an early example. NEC testbed~\cite{nec-testbed} is another. 
NSF PAWR platforms~\cite{pawr} also broadly fall under this category, though each with a different focus (e.g., mmWave in COSMOS~\cite{cosmos}, UAVs for AERPAW~\cite{aerpaw}, rural applications for ARA~\cite{ara}).
Of these, Powder~\cite{powder} features an 8-node outdoor testbed with SDRs deployed across the campus of University of Utah. 
Similar SDR based Open RAN testbeds exist outside the US, including the Open Ireland testbed~\cite{open-ireland}, testbeds in Romania and Norway~\cite{martian25} and various testbeds under Fed4Fire and follow on projects~\cite{fed4fire}.
Note that these SDR based testbeds do not fully support the O-RAN fronthaul interface between DU and RU based on the 3GPP 7.2x split that puts High-PHY on the DU side and Low-PHY on the RU side~\cite{polese2023understanding}. 
As such, they are only partially compliant with the O-RAN standard~\cite{martian25} and consequently also are not compatible with O-RAN compliant commercial virtualized RAN (vRAN) solutions. 
Colosseum facility at Northeastern University~\cite{colosseum} is also based on SDRs but relies instead on hardware based channel emulation, and so can support experimentation based on replaying real-world channel measurement traces.

In contrast to the above, other testbeds have emerged recently that use commercial RUs and are fully O-RAN compliant. 
The testbed at Microsoft Research, Cambridge~\cite{microsoft-testbed} is a building scale indoor deployment with 4 O-RAN compliant Foxconn RPQN-7800 RUs per floor across 5 floors, and uses an Intel FlexRAN based Layer 1 (L1), i.e., PHY~\cite{flexran}. 
X5G testbed at Northeastern is also an indoor testbed that consists of 8 Foxconn RPQN7801 RUs but instead uses NVDIA ARC based GPU accelerated L1.
Recently, a nation wide Open RAN testbed deployment effort was reported from Brazil~\cite{brasil}.
While this testbed is large scale in its geographical span covering multiple interconnected sites across Brazil, each site has a small scale deployment with around handful of commercial RUs (from Foxconn, Benetel and VVDN), mostly indoor. 
The FCT O-RAN testbed at Singapore Institute of Technology~\cite{sit-fct-oran} mixes indoor (3 Foxconn RPQN780) and outdoor (2 Benetel RAN650) RUs with a FlexRAN L1 and Radisys CU / DU.
Our \testbed, while broadly in the same category as the above three testbeds, is deployed outdoors at a campus scale in a dense urban outdoor environment. 
It is currently based on Intel FlexRAN based L1 but is expected to also feature GPU accelerated L1 in the near future. 
Table~\ref{tab:testbed-comparison} summarizes the comparison of \testbed~ with representative existing testbeds.

\section{Network Planning}
\label{sec:planning}

\subsection{Radio Placement and Spectrum}
The goal of the planning process was to find a suitable placement of the radios that would provide both \emph{blanket coverage} across the entire campus and surrounding areas, as well as \emph{sufficient capacity} for known high-footfall areas where we expected larger numbers of users to congregate or traverse on a regular basis. We also preferred to over-provision rather than the opposite, as over-provisioning allows us to slice the network on a radio basis; for example one slice can serve our production users while others can be used for experimental and integration purposes.

An additional constraint was that we did not have unlimited flexibility for where we could deploy the RUs. 
Rather, we had to work within certain constraints placed by the heritage status of several of the university buildings and the feasibility of installing new fiber to buildings to support the O-RAN fronthaul requirements. 
This exercise resulted in identifying six university building sites where we could place the RUs. 

From a spectrum perspective, the only viable option was to use the shared access spectrum made available through Ofcom (the UK communications regulator) for private 5G networks~\cite{ofcom2025sharedaccess}. 
Specifically, there are 4 potential shared access spectrum bands to choose from -- 1800 MHz, 2300 MHz, 3.8-4.2 GHz and 26 GHz.
Of these, very limited amount of spectrum was available to access for 1800 MHz and 2300 MHz bands -- 6.6 MHz and 30 MHz, respectively.
The 2300 MHz band is also restricted to indoor use. 
On the other hand, 26 GHz band provided abundant amount of spectrum -- 3.05 GHz of spectrum -- but would have required 100s of radios to be deployed to achieve campus wide coverage. 

The 3.8-4.2 GHz (n77) band turned out to be a suitable middle ground with up to 390 MHz of spectrum and so we opted for that. 
Within this band, we had two different types of licenses to choose from with different power level restrictions but both allowed channel bandwidths up to 100 MHz.
The `low-power' license allowed power levels up to 21 dBm / 5 MHz for carriers > 20 MHz (EIRP per cell) for multiple cell sites within a 50m radius, whereas the `medium-power' alternative required per-site license while allowing power levels up to 36 dBm / 5 MHz for carriers > 20 MHz.

\begin{figure*}[t]
    \centering
    \includegraphics[width=\linewidth]{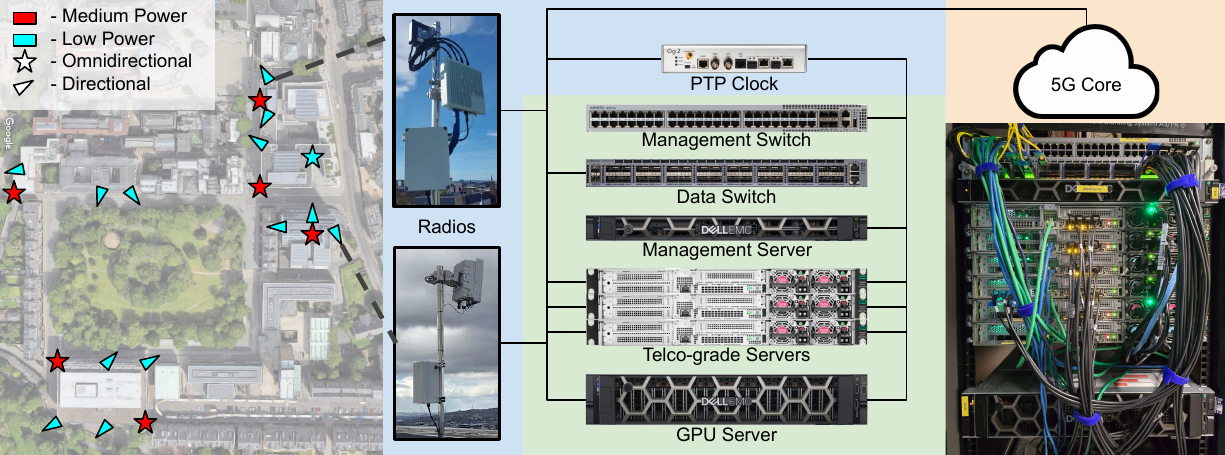}
    \caption{\testbed~physical radio network plan (left) and network architecture (right).}
    \label{fig:map-and-architecture}
\end{figure*}

Given the above constraints, we leveraged our knowledge of the campus environment and footfall in different areas to come up with a radio network plan that would achieve blanket coverage across campus, while ensuring sufficient capacity in all areas. Moreover, we optimized for reducing licensing cost where possible, by using medium-power licenses to reach most of the campus, with the remaining areas served by low-power licenses.
This plan consisted of 20 RUs distributed across 6 building sites on campus along with the choice of power level, antenna type (omni or directional) and orientation for each of them.
Fig.~\ref{fig:map-and-architecture} (left) illustrates this plan overlaid on the campus map. 

\subsection{Pre-Deployment Coverage Map Estimation}\label{sec:pre-deploy}

Here we estimate the radio signal coverage map with our above outlined radio network plan. 
A coverage map \cite{sionnaRT2025} associates signal quality metrics to each discrete bin on a 2D plane, enabling mobile operators to evaluate signal reception at arbitrary locations. 
We mainly focus on the received signal strength indicator (RSSI) metric. 
To achieve a proper trade-off between coverage map resolution and computational complexity, the estimated coverage maps are obtained at the $1$m$\times$$1$m bin granularity within the campus area.

For the coverage map estimation, we employ the Sionna ray tracing (RT) based tool (SionnaRT)~\cite{sionnaRT2025} because it supports accurate ray-tracing-based channel modeling, customized antenna pattern designs, is fully open-source and also GPU-accelerated.
Before using SionnaRT, we first create a real-world campus scene model consisting of three types of objects, i.e., buildings, foliage, and ground (the terrain) with Blender 3.6~\cite{blender3}.
Specifically, we first import the footprints of buildings on campus from OpenStreetMap~\cite{openstreetmap2025} into Blender. 
The foliage is then manually modeled by 3D cuboids, and the terrain information is based on the public LiDAR point cloud data for Scotland~\cite{scotlandLiDAR2024}. 
Each object in Blender is constituted by a mesh of triangles that is acceptable to SionnaRT. 
Furthermore, the radio-frequency materials with the parameters from the ITU standards~\cite{itur2040_2015} are assigned to each object.

With the above scene model in hand, we then setup the ray tracer in SionnaRT as follows. 
The maximum number of interactions (including reflection, diffraction, and scattering) between a ray and a scene object is set to $5$. 
In total, $2\times10^{6}$ random rays are traced. 
The $4$-port omni-directional array pattern and the $4$-port directional array pattern are carefully customized based on the respective antenna data sheets. 
The receiver antenna (reflecting the mobile device) is configured as a single isotropic antenna.

\begin{figure}[t]
    \centering
    \subfigure{
    \includegraphics[width=0.48\linewidth]{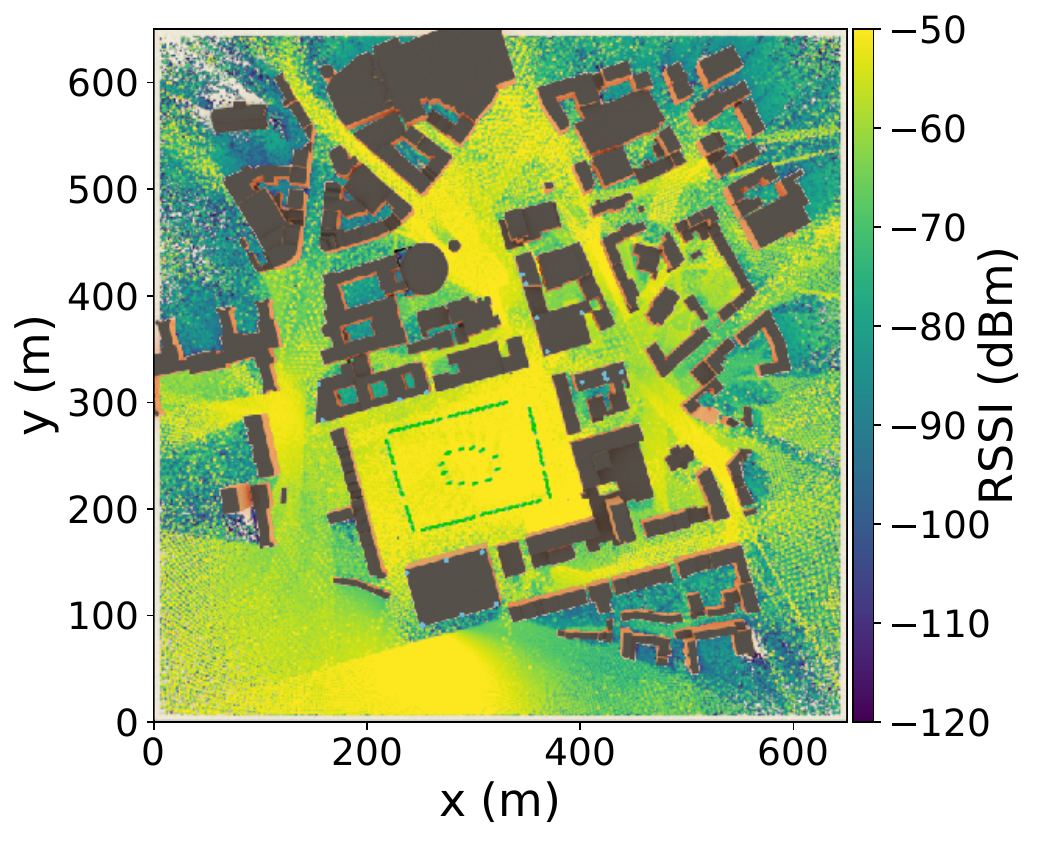}
        \label{subfig:rss_all}
    }
    \hspace{-10pt}
    \subfigure{
        \includegraphics[width=0.48\linewidth]{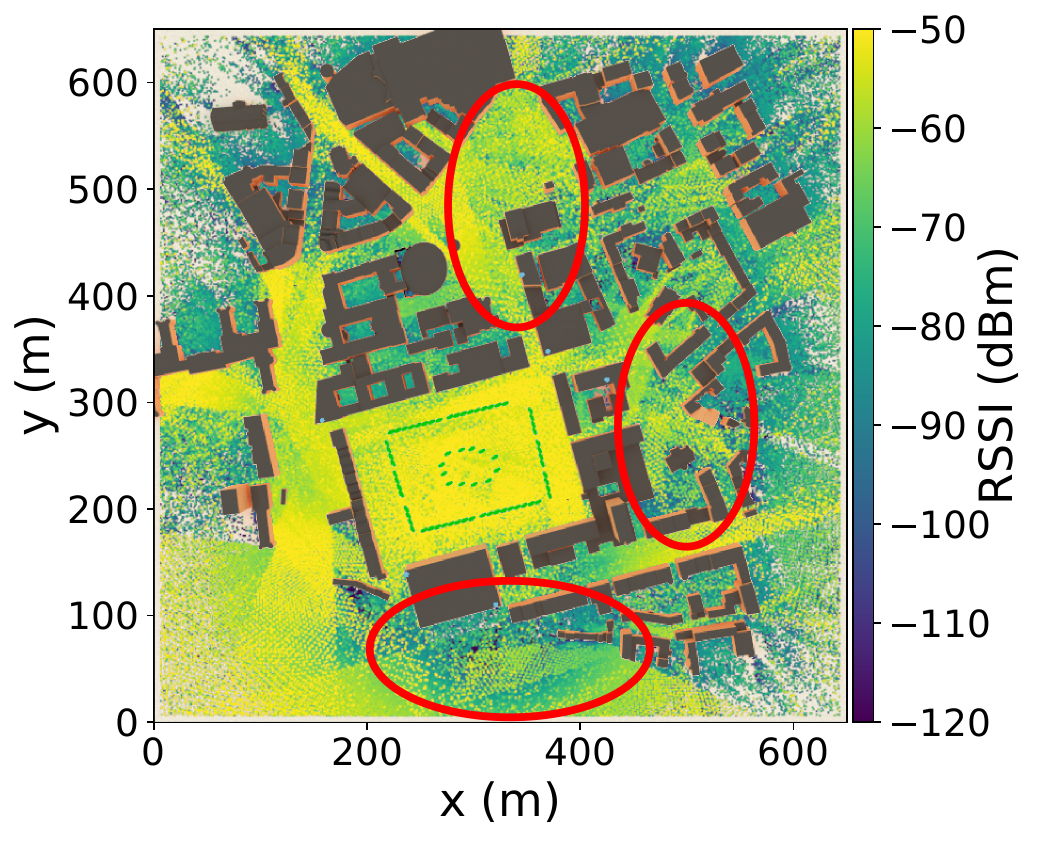}
        \label{subfig:rss_medium}
    }
    \vspace{-5pt}
    \caption{Estimated RSSI coverage maps for two different cases: (a) all radios switched on; (b) only the medium-power radios (with omni-directional antennas) switched on.}
    \vspace{-10pt}\label{fig:rss}
\end{figure}

We now estimate the RSSI coverage maps on the receiver plane at $1$m$\times$$1$m bin granularity for three distinct cases: 1) switching on all the radios; 2) only switching on medium-power radios; and 3) only switching on low-power radios. 
Each of those coverage maps is obtained using the highest RSSI among all the transmitters computed at each bin.
As shown in Fig.~\ref{subfig:rss_all}, when all the 20 radios are switched on, the estimated coverage map reflects blanket coverage across the campus (between $-65$ and $-50$~dBm), with particularly strong coverage in the center of the campus area. 
When only medium-power radios are activated, most regions across the campus still experience satisfactory signal coverage (above $-80$~dBm) as illustrated in Figure~\ref{subfig:rss_medium}. 
But certain areas highlighted by red circles show significant coverage deterioration compared with the scenario with all radios active. 
This result validates our radio network planning strategy, emphasizing the essential role of low-power directional radios in significantly enhancing signal quality in certain parts of the campus. 
\vspace{-2pt}

\section{Testbed Design and Deployment}
\label{sec:design}

\subsection{Architecture and Design Choices}
\label{sec:design-architecture}

From a testbed design perspective, besides the number of radios and their placement covered in the previous section, we have few further design decisions to make: (1) choosing an O-RAN deployment scenario; (2) fronthaul design; (3) handling time synchronization.  
For (1), we opt for the O-RAN deployment scenario ``A''~\cite{oran-deployment} where CU, DU and near real-time (RT) RIC are collocated at an edge cloud while RUs are realized as physical network functions that communicate with DU over the O-RAN fronthaul. 

Regarding (2), we initially considered leveraging existing university fiber infrastructure across the campus. 
But O-RAN fronthaul has stringent latency and bandwidth requirements. 
Specifically, fronthaul latency must be below 100$\mu$s whereas multiple Gbps of fronthaul bandwidth per RU is needed -- the exact breakdown between upstream/downstream dependent on the TDD configuration used~\cite{keysight-fronthaul,o-ran-latencies}. 
As we found this to be infeasible while sharing our university fiber infrastructure, we had to go with custom fiber installation between the edge cloud location and the RUs. 

For (3), for O-RAN to reliably function, RUs must be time synchronized at a fine-grained level (sub-microsecond level) with the functions hosted at the edge cloud, particularly the DU~\cite{cusync}. 
This would only be feasible to achieve with the Precision Time Protocol (PTP)~\cite{ptp-book}.
PTP operates in a master-slave setup, where the master is responsible for serving as the timing source for the slaves and master itself uses a highly-accurate source like GPS as reference. 
In our deployment, we had two options for handling time synchronization: (i) to have a single PTP master for the whole network hosted at the edge cloud; (ii) have an independent PTP master at each RU and the edge cloud. 
Given our decision to have custom fiber installed for the fronthaul, (i) was sufficient to ensure network-wide precise time synchronization. 
Specifically, we used Viavi Qg 2~\cite{viavis-ptp} as the PTP clock in our testbed. 

Our \testbed~testbed architecture following the above design decisions is illustrated in Fig.~\ref{fig:map-and-architecture} (right). 
It consists of eight key components: the radio units (RUs); two different network switches -- one for data traffic, another for management traffic; a computing cluster consisting of a set of telco-grade servers to run the vRAN NFs; a GPU node to process RAN related AI workloads (e.g., AI based Apps); a management node to tie the cluster together; a PTP clock for precise time synchronization; and a cloud-based 5G core.

\subsection{Testbed Components}
\label{sec:design-hardware}

\mypar{Radios}
\label{sec:design-hardware-radios}
We selected two O-RAN compliant RUs from two different vendors: the Benetel RAN650~\cite{benetel-ran650} and the VVDN Mid-Power RU (MPRU)~\cite{vvdn-mpru}. 
These were the only radios we found that meet our criteria: O-RAN 7.2x split compatibility, work in our desired n77 spectrum band, support for 100MHz and 4x4 MIMO, support from both commercial and open source vRANs, and outdoor-rated. 
Our deployment specifically used 12 Benetel and 8 VVDN RUs. 
Note that SDR option is not viable for our outdoor deployment as it did not readily support the O-RAN fronthaul. 

\mypar{Servers}
\label{sec:design-hardware-servers}
The chosen telco-grade servers were the HP DL110 G10 Plus Telco servers, selected due to their track-record of supporting O-RAN workloads \cite{microsoft-testbed}. Each server was fitted with an Intel ACC100 vRAN Accelerator card to provide hardware-acceleration for the High-PHY related signal processing required in the O-RAN DU; these accelerators enabled flexibility through their wide support by commercial RANs.
To ensure adequate network capacity and redundancy, each server was fitted with dual quad-port 25Gbps network cards, giving a total capacity of 200Gbps per server. 
Our current deployment consists of six of these telco servers, the number carefully chosen to meet the processing needs across 20 RUs.

The management and GPU servers were more straightforward: the management server could be any machine capable of acting as the controller in a six-node cluster, whereas the GPU server needed only to support modern graphics cards and high-speed network interconnect to the telco nodes.

\mypar{Switches}
\label{sec:design-hardware-switches}
Our architecture uses two network switches. 
The data switch is the key one that connects the radios to the servers. 
Due to the demanding fronthaul bandwidth and latency requirements, this switch must support 10 Gbps links from each RU, and eight 25 Gbps links from each server. 
To fulfill this requirement, we chose the Arista 7050X3 switch.
The second switch is a management switch, used for keeping management and data traffic separate. 
Unlike the data switch, there are no strict throughput or latency requirements, so we opted for a lower spec Arista 7010X.

\mypar{Platform}
\label{sec:design-platform}
The role of the platform is to provide a software layer atop the hardware, for all the upper layers to run on. 
The platform is a key enabler of flexibility, providing a base for running a diverse set of possible RANs, RICs and Apps.
For the base layer, i.e., the operating system (OS), we chose Ubuntu~\cite{ubuntu}, optimizing it to support real-time processing -- a necessity for the RAN functions. 
Atop the OS, Ansible~\cite{ansible} was used to provision basic utilities such as networking, some drivers, and the Kubernetes orchestrator~\cite{k8s}. 
Kubernetes is the main component of the platform. 
All of the workloads intended to be deployed (RANs, RICs and Apps) were containerized and deployed atop it. 
To manage this orchestration, we made extensive use of Helm charts~\cite{helm}, which allowed `one-click' deployment of a specified RAN setup.

\mypar{Mobile Network Functions}
\label{sec:design-vnfs}
Our platform accommodates a variety of mobile network functions, emphasizing flexibility and interoperability in line with the spirit of Open RAN. 
The platform is designed to seamlessly integrate any O-RAN-compliant vRAN function.
We have extensively tested open-source vRAN solutions, specifically srsRAN~\cite{srsran} and OpenAirInterface (OAI)~\cite{oai}. 
Both srsRAN and OAI can be flexibly deployed either as monolithic network functions, or separated into distinct CU and DU components.
We have also assessed the feasibility of deploying commercial vRAN solutions (e.g., from CapGemini).
In general, with our servers equipped with Intel ACC100 accelerator cards~\cite{acc100}, our infrastructure is capable of running any Intel FlexRAN~\cite{flexran} based vRAN implementation from vendors like CapGemini, Mavenir, Radisys, Parallel Wireless and Rakuten. 

We use srsRAN as the default vRAN software in our deployment and use a split CU-DU architecture with a single CU mapped to 20 DUs (one per RU). 
This setup eases mobility management, specifically enabling intra-CU inter-DU handovers and ensuring seamless user transitions between cells.
For the core network, we currently use an Open5GS~\cite{open5gs} based cloud-hosted solution. 
Our testbed supports access from commodity devices through custom eSIM and physical SIM cards.

\mypar{RICs and Apps}
\label{sec:design-ricapps}
Our platform supports the deployment of standard O-RAN E2~\cite{kpm} based RICs, enabling real-time network optimization through intelligent control loops via xApps. 
Specifically, we have successfully integrated O-RAN SC RIC~\cite{oran-sc} and FlexRIC~\cite{flexric} along with standard applications such as the KPIMON xApp, facilitating basic monitoring and control functionalities.
Given the rising interest in non-standard RICs, which offer more granular control over RAN operations, we have additionally integrated two research RIC solutions: EdgeRIC~\cite{edgeric} and Microsoft Janus/jbpf~\cite{janus,jbpf}.
Our testbed also features advanced AI/ML based Open RAN Apps (e.g., SpotLight~\cite{spotlight}).

\subsection{RAN Configuration}
RAN systems expose hundreds of configurable parameters~\cite{srs_ref, oai_ref} for which vendors typically ship default values for each. 
However, certain parameters must be explicitly updated for the network to function at all. 
These critical settings depend on the specific RU, DU, platform (including OS and server hardware), and the deployment topology.
In our deployment, the initial setup focused on key platform-specific parameters such as network interface bindings, IP addressing, MCC/MNC values, physical cell IDs, and PCI device mappings for DPDK. 
These were combined with DU and RU vendor-specific configurations that were automated using Kubernetes and Helm, streamlining repeatable deployments across multiple nodes. 

However, we found that not all vendor defaults were viable in our environment, and several needed to be manually tuned or overridden.
For instance, we deployed 20 radios overall from two different RU vendors, VVDN and Benetel, each supporting different feature subsets. 
VVDN RUs, for example, support only a fixed TDD configuration and short PRACH format, while Benetel supports seven TDD configurations along with both short and long PRACH formats. 
Additionally, each RU vendor specifies its own fronthaul delay profile~\cite{timing}, but these defaults often do not reflect the real-world deployments. 
In one case, migrating a VVDN RU from a lab testbed using short direct-attach-copper cables to a rooftop site over fiber introduced additional propagation delay, pushing eCPRI timestamps outside the DU’s expected window. 
This resulted in packet loss until we manually adjusted the delay profile to align with the physical link characteristics.

With 20 radios deployed across our campus, most locations observe coverage from 2–3 cells simultaneously. 
Such a deployment density required careful PRACH planning~\cite{9005136}. 
We allocated a unique $prach\_root\_sequence\_index$ to each cell site to prevent random access (RACH) collisions or failures.  
These configurations were sufficient to bring up a stable and functional RAN across our deployment. 
However, they are not yet fully optimized. 
Many additional parameters require further tuning and per-site calibration to reach optimal performance.

With our current deployment and default TDD configuration, we have achieved max downlink (uplink) throughput of 800 Mbps (around 80Mbps).
This is lower than the absolute maximum possible throughput of 1.4Gbps with 100MHz and 4x4 MIMO, suggesting room for optimization through configuration parameter tuning. 
To put these numbers in context, what we already achieve is more than 3.5 times the max downlink throughput measured with public 5G mobile networks deployed in our campus area.

\section{Measurement based Coverage Analysis}
\label{sec:coverage-analysis}

In this section, we present measurement data from \testbed~collected through various walking paths across the campus. 
Since measurements cannot cover every possible location within the target area, interpolation is required to generate a complete coverage map. 
To determine an efficient interpolation method, we first evaluate candidate schemes using simulated coverage data produced by SionnaRT (\S\ref{sec:pre-deploy}). 
Then, we introduce the collected real-world measurement dataset, and the selected interpolation approach is applied to the measured data to create complete coverage maps.

\subsection{Comparison of Interpolation Methods}\label{subsec:interp}
The goal of spatial interpolation is to obtain the values at the unsampled locations. Let $\hat{Z}(\mathbf{s}_0)$ be the estimate at an unsampled location $\mathbf{s}_0$
from observations $\{Z(\mathbf{s}_i)\}_{i=1}^{N}$ where $\mathbf{s}_i$ denotes an observed location. To determine the best interpolation method for the coverage map, we group approaches into four distinct families (outlined in the Appendix~\ref{interpolation-methods}) based on how they construct $\hat{Z}(\mathbf{s}_0)$ and benchmark one or two representative algorithms from each family:

\begin{tightlist}
\item \textbf{Deterministic}: \emph{Inverse Distance Weighting} (IDW) and \emph{Radial Basis Function} (RBF).
\item \textbf{Geostatistical}: \emph{Ordinary Kriging} (OK).
\item \textbf{Machine learning based}: \emph{Random Forest} (RF) and \emph{XGBoost}.
\item \textbf{Physics aware}: \emph{Model-based radio interpolation} (MRI)~\cite{MRI}.
\end{tightlist}

% use sionna to evaluate different interpolation methods; sionna dataset setup (split ratio, outlier processing)
% results in tables/figures
\begin{table}[!t]                % only allow Top placement, ‘!’ to override defaults
  \centering
  \caption{Comparison of various interpolation methods.}
  \label{tab:interp_performance}
  \begin{tabular}{@{} ll *{3}{>{\centering\arraybackslash}p{1.5cm}} @{}}
    \toprule
    \multirow{2}{*}{\textbf{Category}} 
      & \multirow{2}{*}{\textbf{Method}}
      & \multicolumn{3}{c}{\textbf{Metrics on five folds (Avg. $\pm$ Std.)}} \\
    \cmidrule(lr){3-5}
      & 
      & \textbf{RMSE} & \textbf{NMSE} & \textbf{MAPE (\%)} \\
    \midrule
    \multirow{3}{*}{Deterministic}
      & IDW    & $8.21\pm0.14$ & $0.24\pm0.01$  & $7.06\pm0.15$ \\
      & RBF   & $\mathbf{7.25\pm0.14}$ & $0.19\pm0.01$ & $\mathbf{5.85\pm0.10}$ \\
    \midrule
    Stochastic
      & OK     &$7.37\pm0.10$  &$\mathbf{0.19\pm0.00}$  &$5.95\pm0.04$  \\
    \midrule
    ML-based
      & RF    &$7.46\pm0.06$  &$0.20\pm0.00$  &$6.03\pm0.02$  \\
      & XGBoost    & $7.77\pm0.09$ & $0.21\pm0.00$ & $6.44\pm0.05$ \\
    \midrule
        Physics-aware
      & MRI  &$7.40\pm 0.15$  & $\mathbf{0.19 \pm 0.00}$ & $6.05 \pm 0.06$ \\
    \bottomrule
  \end{tabular}
\end{table}

To evaluate these algorithms, we perform 5-fold cross-validation on the RSSI dataset generated by SionnaRT, using $80\%$ of the data for training and the remaining $20\%$ for testing in each fold. 
The average and standard deviation of three metrics, i.e., Root Mean Square Error (RMSE), Normalized Mean Square Error (NMSE), and Mean Absolute Percentage Error (MAPE), are computed across folds, and reported in Table~\ref{tab:interp_performance}. 
These results show that RBF achieves the best RMSE and MAPE performance, while Ordinary Kriging and MRI are marginally better in terms of NMSE and its standard deviation. 
So, we select RBF to interpolate measurement based coverage maps, specifically with a multi-quadratic kernel (shape parameter $\epsilon=1$ and smoothing parameter $\delta=0.1$).

\subsection{Measurement Based Coverage Maps}
\label{sec:coverage-measurement}
% hardware/software (timetag, gps locations)
We conducted a ~15 Km walking test across the campus to collect measurements using a commodity 5G smartphone, specifically Nothing Phone 2~\cite{np2}, including GPS locations and UE reported metrics (RSRP, RSRQ and SINR). 
Around 11K samples in the form of [GPS location, reported metrics] are collected. 
To align the metrics obtained from Sionna and field measurements, we compute RSSI from the measured data based on the following relationship~\cite{3gpp.38.213}:
\begin{equation}
\mathrm{RSSI}_{\mathrm{dBm}}=\mathrm{RSRP}_{\mathrm{dBm}}+10\log_{10}N -\mathrm{RSRQ}_{\mathrm{dB}},
\end{equation}
where $N$ is the number of physical resource blocks (PRBs) in the measurement bandwidth configured in SS/PBCH block ($N=20$ for our testbed).

\begin{figure}[t]
    \centering
    \subfigure{
        \includegraphics[width=0.50\linewidth]{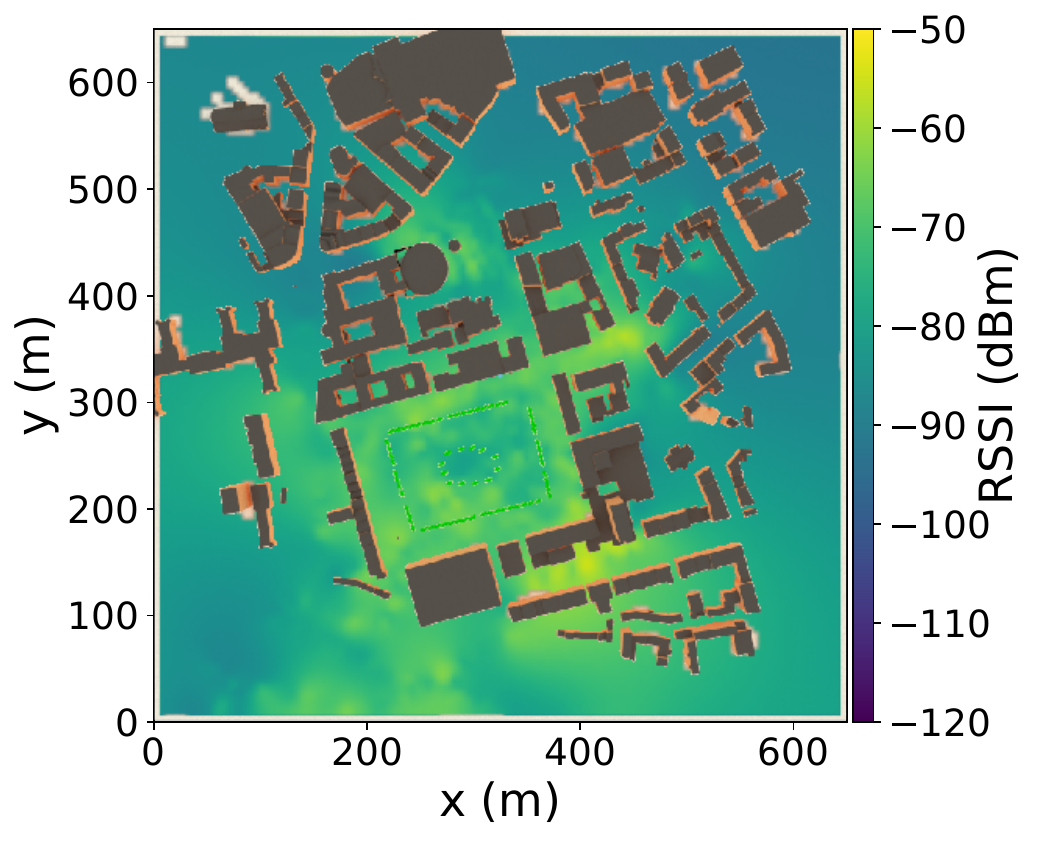}
        \label{subfig:rss_after_interp}
    }
    \hspace{-10pt}
    \subfigure{
        \includegraphics[width=0.48\linewidth]{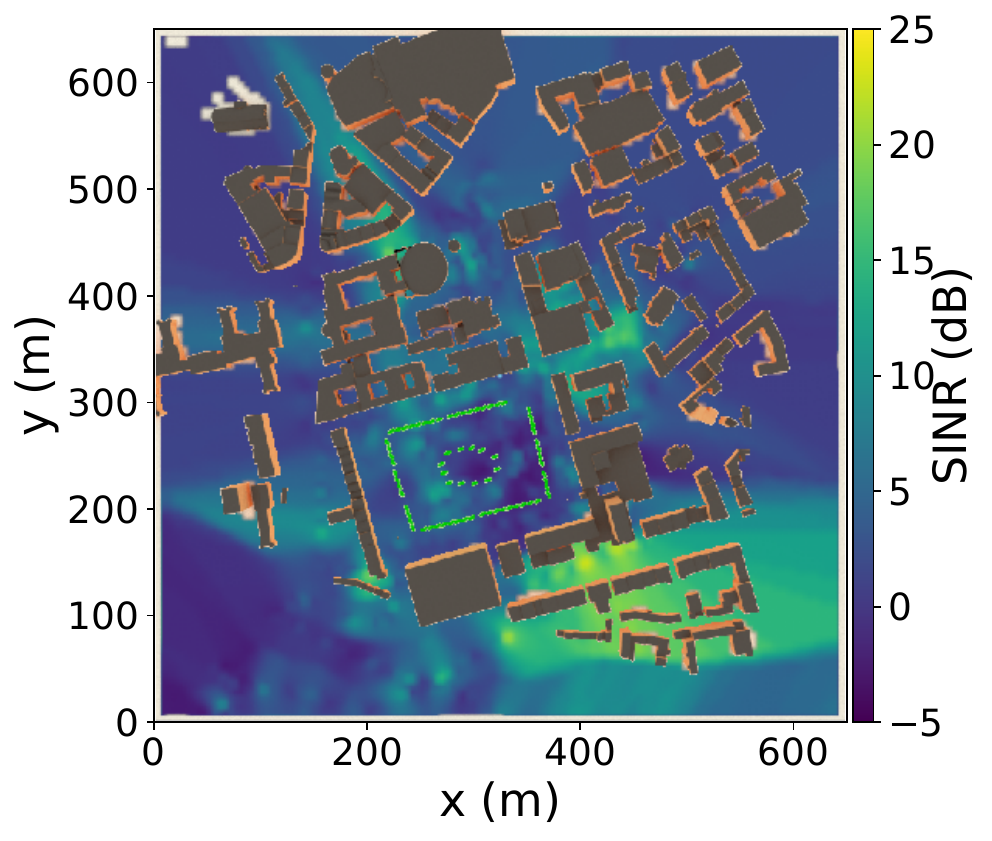}
        \label{subfig:sinr_after_interp}
    }
    \vspace{-8pt}
    \caption{Measurement based RSSI coverage (left) and SINR (right) maps.}
    \vspace{-10pt}\label{fig:meas}
\end{figure}

The measurement based RSSI coverage map via RBF interpolation is shown in Fig.~\ref{subfig:rss_after_interp}.
As with the earlier ray-tracing based estimated coverage map in Fig.~\ref{fig:rss}, measured map also confirms blanket coverage across the campus area. 

To analyze the tradeoff between coverage and interference, we also interpolate the SINR metric to obtain the SINR based coverage map, as shown in Fig.~\ref{subfig:sinr_after_interp}. 
While our testbed delivers high signal strength in the central campus area, it also leads to severe interference (resulting in low SINR levels between -5 dB and 5 dB) when all radios are operating at full power. 
This highlights the critical trade-off between signal coverage and interference in dense deployments, necessitating careful interference management to optimize network performance.

\section{Lessons Learned}
\label{sec:lessons-learned}

\mypar{Cellular is attractive for campus wide coverage}
Our post-deployment measurement based analysis of \testbed~coverage revealed that it was remarkably resilient when compared to the existing WiFi network on campus. In scenarios where the WiFi network would drop out, our testbed (with the equivalent signal strength) worked well. This demonstrates the advantages of deploying cellular over WiFi, aligned with observation made in other recent work~\cite{foukas-hotmobile25}. 
This is particularly the case in large areas such as open spaces in university campuses, which would otherwise require a large number of WiFi access points to be deployed~\cite{seda-dyspan25}.

\mypar{First, make the platform real-time}
Ensuring real-time behavior is essential to run latency-sensitive DU workloads. Installing a real-time kernel is only the first step. 
Meaningful performance requires careful tuning of both hardware and software. This includes configuring BIOS settings to disable power-saving features, isolating CPU cores for RAN functions, removing interrupts from those cores, and disabling low-power C-states. We used tools like tuned~\cite{tuned} along with x86 processor-specific guidelines from Intel~\cite{intel-tuned} and AMD~\cite{amd-tuned} to apply these changes. Without these steps, even optimized RAN stacks showed inconsistent performance under load.

\mypar{No two vendors are the same}
We observed considerable variation between vendors, both in vRAN NFs (DU, CU) and O-RUs. Features like TDD configurations, PRACH support, and delay handling differ widely. In many cases, vendor defaults were incompatible with our environment, and given the early maturity of open RAN ecosystems, support teams were not always able to help. These differences meant we often had to debug and adapt configurations ourselves. Seamless multi-vendor interoperability remains a challenge, requiring significant deployment-specific tuning.

\mypar{Performance vs. stability trade-off}
Configuring RAN for maximum performance can introduce system-level issues. 
Aggressively prioritizing RAN threads may yield higher throughput but risks starving kernel processes or other workloads of shared resources, such as CPU caches. 
On systems using a real-time kernel, this can make the platform unresponsive, as even essential kernel threads may be blocked. 
In our experience, slightly trading off from peak performance in favor of system stability yields more reliable and maintainable deployments.

\mypar{Deployment costs are insightful}
Examining costs of deployment reveals that civil engineering costs (fiber, electrical and antenna installation) make up a substantial fraction (at least a third) of the overall CAPEX costs. 
However, overall deployment costs for a campus scale private 5G network like our \testbed~testbed are surprisingly in the same ballpark as the costs for deploying a single macrocell in a dense urban area~\cite{wisely18}.

\section{Research Opportunities}
\label{sec:research-opportunities}

Our experience deploying the \testbed~testbed brought to the fore several key research opportunities that we outline below.

RAN configuration and optimization is a broad topic for future work. 
Even small changes in RAN configuration parameters can lead to significant behavioral shifts, motivating research on their optimization and dynamic adaptation. 
An illustrative example from our deployment is the impact of tuning the measurement reporting interval on handovers. 
In a controlled walk-around experiment we conducted on campus, we evaluated how different reporting interval settings affect the number of handovers triggered for the same physical route. 
Table~\ref{tab:ho} summarizes the results from this experiment. 

The above example conveys how parameter tuning can influence not only the signaling overhead and mobility behavior but also user experience and system load. 
However, the optimal configuration is often context-dependent, influenced by mobility patterns, network density, and cell overlap. 

\begin{table}[!t]
\centering
\captionsetup{justification=centering}
\caption{Impact of measurement reporting interval on handover frequency.}
\begin{tabular}{|c|c|}

\hline
\textbf{Report Interval (ms)} & \textbf{\#Handovers} \\ \hline
480                  & 133            \\ \hline
1024                 & 39             \\ \hline
5120                 & 23             \\ \hline
\end{tabular}

\label{tab:ho}
\vspace{-8pt}
\end{table}

Systematic exploration of these parameter spaces, especially in conjunction with RIC based closed-loop control, represents a rich area for future research.
Interference management as alluded to in \S\ref{sec:coverage-measurement} and Fig.~\ref{subfig:sinr_after_interp} is another case demanding further work. 
More issues to look at along these lines include: (1) intelligent and dynamic use of spectrum within a private 5G network and across networks; (2) continual optimization of resource and energy usage, which requires a holistic understanding of energy consumption and its interplay with the use of compute and radio resources. 

Complementarily, monitoring and optimizing user experience is another topic with several research opportunities.  
Our measurement effort in \S\ref{sec:coverage-measurement} is a case in point. 
To obtain data for measurement based coverage maps, we conducted a ~15 Km walk test to evaluate RAN behavior across our deployment. 
While effective, this process was very manual and required constant supervision. 
Automating such measurements towards zero-touch management for private 5G and Open RAN presents a strong research opportunity. 
Another significant one concerns optimizing user experience while moving both within the coverage area of a private network as well as at its edges with other networks.

\section{Conclusions}
\label{sec:conclusions}

In this paper, we have captured our journey planning, deploying and measuring \testbed, the first-of-its kind campus scale O-RAN testbed. 
Urban, outdoor and university campus setting that our deployment represents is a perfect blend for innovation on future wireless access in a real-world environment. 
We have reflected on our experience deploying \testbed~and shared the key lessons.
We have also outlined a number of research opportunities that came to the surface through our deployment experience. 

\begin{acks}
This work was mainly supported by a project funded by the UK Department for Science, Innovation and Technology (DSIT). 
The work of T. Wang and M. Marina was also supported by the UKRI/EPSRC grant UKRI860.
\end{acks}
\bibliographystyle{ACM-Reference-Format}
\bibliography{reference}

\appendix

\section{Overview of Interpolation Methods}
\label{interpolation-methods}

\begin{itemize}
  \item \textbf{Deterministic}. The spatial field is assumed deterministic in this category. The interpolator is a linear/non-linear function of known data points. \emph{Inverse Distance Weighting} (IDW) is the canonical linear method that embodies pure distance-decay weighting, while \emph{Radial Basis Function} (RBF) non-linear interpolation adds global smoothness via a radial kernel. We specifically consider a multi-quadratic kernel for RBF.

  \item \textbf{Geostatistical}. The spatial field is modeled by a random process with a specified covariance (variogram) structure. Among the algorithm in this category, \emph{Ordinary Kriging} (OK) is the canonical case with an unknown but constant mean and it serves as the benchmark for kriging variants and Gaussian process regression methods~\cite{zipweave}.

  \item \textbf{Machine-learning-based}. In this category, location coordinates (and any covariates) form a feature vector $\mathbf{x}$, and then a flexible function $\hat{Z}(\mathbf{s}_0)=f(\mathbf{x}_0;\,\theta)$ is learned by minimizing an interpolation loss~\cite{ML_based_interp}, where $\theta$ denotes unknown parameters. We consider two effective ML-based algorithms that are widely used, namely \emph{Random Forest} (RF) and \emph{XGBoost}. Both algorithms scale well to large spatial datasets and require much lighter computation than the neural network based methods.

  \item \textbf{Physics-aware}. Instead of relying solely on data, physics-aware methods incorporate physical laws/constraints into the interpolation process. A representative method in this category is \emph{Model-based radio interpolation} (MRI)~\cite{MRI}, which assumes a log-distance path-loss model for each transmitter and fits its parameters using linear regression.
\end{itemize}

\end{document}